\documentclass[12pt,reqno]{amsart}
\usepackage{amsthm}
\usepackage{amsmath}
\usepackage{latexsym}
\usepackage{amsfonts}
\usepackage{amssymb}
\usepackage{color}
\usepackage{bbm,dsfont}
\usepackage{graphicx}
\usepackage{hyperref}
\usepackage{enumerate}
\usepackage{comment}
\usepackage{caption}
\usepackage{subcaption}
\usepackage{amsaddr}

\theoremstyle{definition}

\newcommand{\real}{\mathbb{R}} 

\newcommand{\half}{\tfrac{1}{2}} 
\newcommand{\mo}[1]{\left| #1 \right|} 

\newcommand{\hi}{\mathcal{H}} 
\newcommand{\lh}{\mathcal{L(H)}} 

\newcommand{\ip}[2]{\left\langle\,#1\,|\,#2\,\right\rangle} 
\newcommand{\ket}[1]{|#1\rangle} 
\newcommand{\kb}[2]{|#1\rangle\langle#2|} 
\newcommand{\no}[1]{\left\|#1\right\|} 
\newcommand{\tr}[1]{{\rm tr}\left[#1\right]} 
\newcommand{\ptr}[2]{\textrm{tr}_{#1}[#2]} 

\newcommand{\id}{\mathbbm{1}} 

\renewcommand{\rho}{\varrho}

\newcommand{\va}{\mathbf{a}} 
\newcommand{\vb}{\mathbf{b}} 
\newcommand{\vx}{\mathbf{x}} 
\newcommand{\vy}{\mathbf{y}} 
\newcommand{\vz}{\mathbf{z}} 
\newcommand{\vsigma}{\boldsymbol{\sigma}} 

\newcommand{\Ao}{\mathsf{A}}
\newcommand{\Bo}{\mathsf{B}}
\newcommand{\Co}{\mathsf{C}}
\newcommand{\Go}{\mathsf{G}}
\newcommand{\Mo}{\mathsf{M}}
\newcommand{\Io}{\mathsf{I}}

\newcommand{\en}{\mathcal{E}} 

\newcommand{\Ii}{\mathcal{I}} 

\newcommand{\pg}{P_{\mathrm{guess}}} 

\begin{document}\setlength{\arraycolsep}{2pt}

\title[]{Quantum Incompatibility from the viewpoint of entanglement theory}

\author{Teiko Heinosaari}
\address{QTF Centre of Excellence, Turku Centre for Quantum Physics, Department of Physics and Astronomy, University of Turku, FI-20014 Turku, Finland}

\begin{abstract}
In this essay I discuss certain analogies between quantum incompatibility and entanglement.
It is an expanded version of my talk presented at \emph{Mathematical Foundations of Quantum Mechanics in memoriam Paul Busch}, held in York in June 2019.
\end{abstract}

\maketitle

\section{Introduction}

The basics of entanglement theory belong to any first course on quantum information, while incompatibility is not necessarily covered at all, mentioned just briefly, or perhaps falsely identified with non-commutativity.
For this reason, in the following I will use some elementary well-known facts about entanglement to guide the presentation on quantum incompatibility through some analogies.
I hope this could be helpful for a reader who has not encountered with incompatibility so much.
For a more expert reader the highlighted analogy between entanglement and incompatibility may give thoughts about additional, yet unexplored, connections.
My intention has been to keep the presentation light, simple and concrete, therefore I'm using only elementary mathematical machinery and not e.g. the resource theoretic framework where some of the discussed features could be treated in a unified manner.

Personally, the analogy between entanglement and incompatibility became visible to me when I visited Paul in Perimeter Institute in February 2007 and we were continuously discussing incompatibility of quantum observables.
Before that, I had studied incompatibility of certain pairs of quantum observables, such as noisy position and noisy momentum, and I had also worked with Paul on those things.
During that visit we started to talk with Paul about a goal of having a `theory of incompatibility', something similar to entanglement theory.
This basically meant that we wanted to study general features of incompatibility, not only some physically interesting examples. 
That is what we then both devoted time to in the following years, together and separately.
We still agreed that physically relevant examples are important to find new mathematical techniques and to demonstrate the domain of incompatibility, so that research line was not forgotten.
But since that visit the general features of incompatibility have been one of my main research interests and the visit had, in fact, longlasting implications on my research.
In all of the subtopics discussed in this paper Paul has either directly worked with me, boosted my research with discussions on those topics, or encouraged me to investigate some questions. 
It is clear to me that I wouldn't have involved myself so deeply into incompatibility questions without his encouragement and support.

The general goal of `incompatibility theory' was made public in one form in an article of Paul and Heinz-J\"urgen Schmidt a couple of years after my visit at the Perimeter Institute; in the introduction of \cite{BuSc10} they wrote that: \emph{The present paper is a contribution to the emerging programme of investigating the structure of the set of observables, which should complement current studies of the dual structure of the set of quantum states.}
That sentence does not mention `incompatibility' so it can refer even to a more general programme, but the article is, in fact, about incompatibility and I believe that our discussions at the Perimeter partly motivated that sentence.
Other two public `announcements' of the goal can been seen as the two workshops devoted solely to incompatibility. 
These were held at the Technical University of Munich in 2013 and at Maria Laach Abbey in 2017.
The workshops were in large part organized by our friends and colleagues, but Paul had an important role for the events to happen.

I need to make two beginning remarks. 
Firstly, this paper is not meant to be a review paper. 
I have chosen rather arbitrarily some topics that have been of my personal interest and there are many more that are not mentioned here at all. 
A more systematic invitation to incompatibility is given in \cite{HeMiZi16}, although many important results have been found after that article appeared.
Secondly, I have written some personal comments (like those above) on the background of some of the research results.
They are subjective memories and someone else can remember the events differently.

\section{Definition of incompatibility}

Entanglement is a property of bipartite (or multipartite) states, while incompatibility is a property of pairs (or collections) of observables.
The first similarity between entanglement and incompatibility is that they are defined by saying what they are not.
Namely, an entangled state is a composite state that is not separable, and a separable state $\omega$ is, by definition, a convex mixture of product states,
\begin{equation}\label{eq:sep}
\omega = \sum_i t_i \ \varrho^A_i \otimes \varrho^B_i \, .
\end{equation}
Similarly, an incompatible collection of observables is a collection that is not compatible, and I will shortly recall the definition of compatibility.
In conclusion, both separability and compatibility have clear operational definitions, but entanglement and incompatibility get their meaning in this negative way.
In the following I will concentrate on pairs of observables, similarly as one would start entanglement theory with bipartite states. 
I make some comments on the general case in Section \ref{sec:multi}.

A quantum observable is mathematically described as a positive operator valued measure (POVM).
For simplicity, in this paper all observables are assumed to have finite number of outcomes. 
An observable $\Ao$ is a hence a map $x \mapsto \Ao(x)$ from a finite set of outcomes $[m]\equiv \{1,\ldots,m\}$ to the algebra $\lh$ of bounded linear operators on a Hilbert space $\hi$, such that each $\Ao(x)$ is a positive operator (i.e. $\ip{\psi}{\Ao(x)\psi}\geq 0$ for all $\psi\in\hi$) and $\sum_x \Ao(x) = \id$.
The physical interpretation is that if the system is prepared in a state $\varrho$ and a measurement of $\Ao$ is performed, then the outcome $x$ is obtained with probability $\tr{\varrho \Ao(x)}$.
Many of the definitions and results are equally valid for finite and countably infinite dimensional Hilbert spaces.
However, I will assume that Hilbert spaces are finite dimensional as this is the usual setting where one learns entanglement theory and I want to keep the discussion simple.

\begin{figure}[t]
    \centering
    \includegraphics[width=0.6\textwidth]{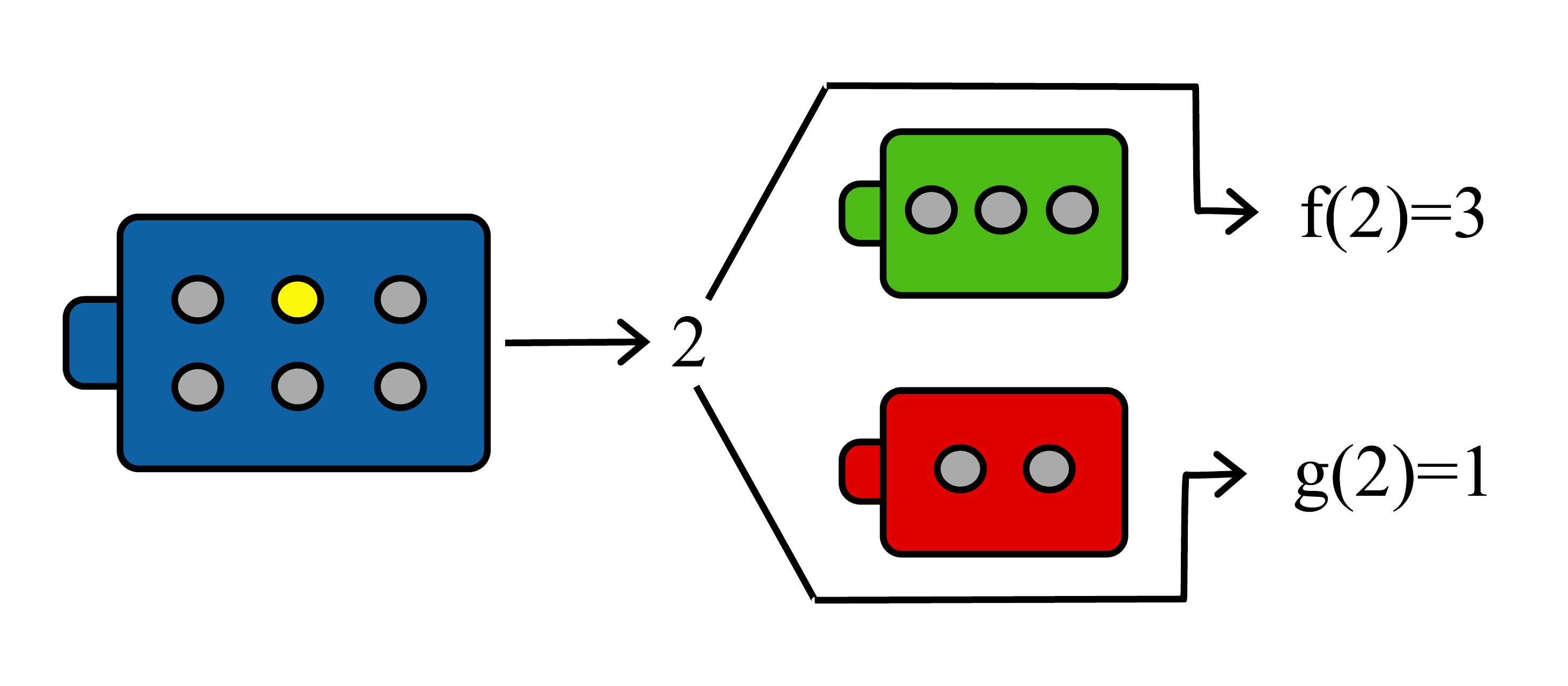}
    \caption{\label{fig:relabeling} Two observables (green and red) are compatible if their measurements can be simulated with a single observable (blue) by relabeling the obtained measurement outcomes.}
\end{figure}

Two observables $\Ao$ and $\Bo$, with outcome sets $[m]$ and $[n]$, are \emph{compatible} if there exists a third observable $\Co$, with an outcome set $[k]$, and functions $f:[k]\to [m]$ and $g:[k]\to[n]$ such that
\begin{equation}\label{eq:relabel}
\Ao(x) = \sum_{z:f(z)=x} \Co(z) \, , \quad \Bo(y) = \sum_{z:g(z)=y} \Co(z) \, .
\end{equation}
The functions $f$ and $g$ are simply relabeling the outcomes of $\Co$.
The intuitive idea is depicted in Fig. \ref{fig:relabeling}.
If $\Ao$ and $\Bo$ are not compatible, then they are \emph{incompatible}.
The power of stating the definitions of compatibility and incompatibility in this way is that they have direct generalizations to any number of observables, even infinite number. 

There is a useful equivalent definition when we are considering incompatibility of finite number of observables.
If $\Ao$ and $\Bo$ are compatible and $\Co,f,g$ are such that \eqref{eq:relabel} holds, then we can define a new observable $\Go$ on the product outcome set $[m] \times [n]$ as
\begin{equation}
\Go(x,y) = \sum_{z:f(z)=x \wedge g(z)=y} \Co(z)
\end{equation}
and then \eqref{eq:relabel} implies that
\begin{equation}\label{eq:joint}
\Ao(x)=\sum_y \Go(x,y) \, , \quad \Bo(y)=\sum_x \Go(x,y) \, .
\end{equation}
This is a special case of \eqref{eq:relabel}, where the relabeling functions are taken to be the projections $f(x,y)=x$ and $g(x,y)=y$.
Any observable $\Go$ satisfying \eqref{eq:joint} is called a \emph{joint observable} of $\Ao$ and $\Bo$.
We conclude that one can take the existence of a joint observable as an equivalent definition of compatibility for two (or finite number of) observables. 
Further, it can be shown that even if relabeling functions would be stochastic, it is still possible to construct a joint observable \cite{AlCaHeTo09}.
This means that by allowing relabeling functions to take different values with some fixed probabilities (hence being Markov kernels), the definition of compatibility is still the same.

\section{The basic case}\label{sec:simple}

When investigating some mathematically defined property, it is very useful to have a simple `test class' of objects where that property is easy to check.
The simplest case in entanglement theory is the composite state space of two qubits.
The entanglement in this class of states has a neat analytic solution: a bipartite qubit state is entangled if and only if its partial transpose has a negative eigenvalue \cite{Peres96,HoHoHo96}.

The simplest class of pairs of observables is the class of two unbiased dichotomic qubit observables. 
This class has turned out to be very useful in testing (in)compatibility conditions, quantifications and many other developments. 
An unbiased dichotomic qubit observable is determined by a vector $\va\in\real^3$, $\no{\va}\leq 1$, via 
\begin{equation}
\Mo^{\va}(\pm) = \half ( \id \pm \va \cdot \vsigma ) \,,
\end{equation}
where $\va \cdot \vsigma = a_1\sigma_1 + a_2\sigma_2 + a_3\sigma_3$ and $\sigma_i$ are the Pauli matrices.
(It is customary and convenient to label the outcomes $\pm$ instead of $1,2$.)
The necessary and sufficient incompatibility condition was found by Paul in his seminal work \cite{Busch86}: $\Mo^{\va}$ and $\Mo^{\vb}$ are compatible if and only if 
\begin{equation}\label{eq:paul}
\no{\va + \vb} + \no{\va - \vb} \leq 2 \, .
\end{equation}
The physical interpretation of \eqref{eq:paul} has been analyzed in detail by Paul and his coworkers in \cite{Busch87,BuSh06,BuBu18}.

A general (i.e. not necessarily unbiased) dichotomic qubit observable has an additional parameter $\alpha\in[0,1]$ and has the form
\begin{equation}
\Mo^{\alpha,\va}(\pm) = \half ( (1\pm\alpha) \id \pm \va \cdot \vsigma ) \,,
\end{equation}
with the parameters satisfying $\mo{\alpha} \leq 1-\no{\va}$.
We did consider this class of observables with Paul in \cite{BuHe08} and found some partial results on their compatibility, but it took some time to digest the ideas and find the final answer.
The necessary and sufficient condition for $\Mo^{\alpha,\va}$ and $\Mo^{\beta,\vb}$ being compatible was found independently (in different but equivalent forms) in \cite{BuSc10,StReHe08,YuLiLiOh10}. 
Paul and I knew that we are working on the same problem, but our approaches and mathematical techniques were so different that we decided to keep the works separated.
We coordinated the arXiv submissions and they appear with the same title and in the same day in February 2008 in arXiv.
Admittedly, the form proven in \cite{YuLiLiOh10} is particularly concise; $\Mo^{\alpha,\va}$ and $\Mo^{\beta,\vb}$ are compatible if and only if
\begin{equation}\label{eq:compatible}
\left( 1-s(\alpha,a)^2 - s(\beta,b)^2 \right) \left( 1- \frac{\alpha^2}{s(\alpha,a)^2} - \frac{\beta^2}{s(\beta,b)^2} \right) \leq (\va \cdot \vb - \alpha \beta)^2 \, .
\end{equation}
where $a=\no{\va}$, $b=\no{\vb}$ and $s$ is a function defined as
\begin{equation}
s(\alpha,a) = \half \left( \sqrt{(1+\alpha)^2 - a^2} + \sqrt{(1-\alpha)^2 - a^2} \right) \, .
\end{equation}
The meaning of $s$ (or more precisely $1-4s^2$) as a sharpness measure was discussed by Paul in \cite{Busch09}.

A simple but important observation of Paul in \cite{Busch86}, preceeding the proof of \eqref{eq:paul}, was that a joint observable of any pair of dichotomic observables (i.e. not only qubit) is determined by a single operator.
In fact, dichotomic observables $\Ao$ and $\Bo$ are compatible if and only if there exists $G\in\lh$ such that the following four operators are positive:
\begin{itemize}
\item[(i)] $G$ 
\item[(ii)] $\Ao(1)-G$
\item[(iii)] $\Bo(1) - G$
\item[(iv)] $\id + G - \Ao(1)-\Bo(1)$
\end{itemize}
The equivalence is straightforward to verify: the operators (i)--(iv) are just the operators $\Go(x,y)$, $x,y=\pm$, for a joint observable $\Go$.
This characterization gives a starting point for any attempt to determine the compatibility of two dichotomic observables belonging to some specific class, e.g. when the operators belong to the von Neumann algebra generated by two projections \cite{HeKiRe14}. 
The characterization further shows that the compatibility of $\Ao$ and $\Bo$ can be decided efficiently numerically:
the problem
\begin{equation}
\inf \{ t \in \real : t\id + G \geq \Ao(1)+\Bo(1)\} \, , 
\end{equation}
subject to the positivity constrains of operators (i)-(iii), is a semidefinite program \cite{WoPeFe09}.

\section{Convexity}\label{sec:convex}

The set of all separable states is a convex subset of all bipartite states. 
This property is a starting point for many theoretical developments.
For instance, one can quantify entanglement of a state $\omega$ as the least upper bound of all numbers $t\in [0,1]$ such that the mixed state $t \omega + (1-t) \tfrac{1}{d_A d_B} \id$ is entangled, or alternatively
$t \omega + (1-t)  \sigma_{\mathrm{sep}}$ is entangled for all separable states $\sigma_{\mathrm{sep}}$.
These and similar ways to quantify entanglement have been developed in great detail \cite{HoHoHoHo09}.

The convexity plays an important role also in investigations of incompatibility.
Let us first recall that the set of observables with a fixed outcome set is a convex set.
The convex mixture of $\Ao$ and $\Ao'$ with a mixing parameter $0\leq t \leq 1$ is defined as
\begin{equation}
(t \Ao + (1-t) \Ao')(x) = t \Ao(x) + (1-t) \Ao'(x) \, .
\end{equation}
In \cite{BuHe08} the following was noted.
For two pairs of observables $(\Ao,\Bo)$ and $(\Ao',\Bo')$ and a mixing parameter $t$, define the mixture as
\begin{equation}
t (\Ao,\Bo) + (1-t) (\Ao',\Bo') = (t\Ao + (1-t) \Ao',t\Bo+(1-t)\Bo') \, .
\end{equation}
Here $\Ao$ and $\Ao'$ are assumed to have same outcome set and same for $\Bo$ and $\Bo'$, but the two outcome sets can be different. 
Importantly, if both pairs $(\Ao,\Bo)$ and $(\Ao',\Bo')$ are compatible, then also their mixture is compatible.
Namely, if $\Go$ is a joint observable of $(\Ao,\Bo)$ and $\Go'$ is a joint observable of $(\Ao',\Bo')$, then it is straightforward to verify from the definition of joint observable that the mixture $t \Go + (1-t) \Go'$ is a joint observable of $t (\Ao,\Bo) + (1-t) (\Ao',\Bo')$.

One possible quantification of incompatibility is the least upper bound of all numbers $t\in [0,1]$ such that the mixture $t (\Ao,\Bo)+(1-t) (\Io_m,\Io_n)$ is incompatible for all trivial observables $\Io_m$ and $\Io_n$.
An $m$-outcome trivial observable $\Io_m$ has the form $\Io_m(x)=p(x) \id$ for some probability vector $p$ of length $m$.
As with the previously mentioned ways to quantify entanglement, one can vary also this quantification in several ways, depending to what kind of pairs one mixes the quantified pair $(\Ao,\Bo)$ in question.
All these kind of quantifications are based on the convexity of the set of compatible pairs of observables.
A recent work \cite{DeFaKa19} reviews several different quantifications while the general convex framework has been clarified in \cite{Haapasalo15}.

Any quantification of incompatibility leads naturally to the question of \emph{maximally incompatible} pairs of observables. 
I remember when I first time started to talk with Paul about this question.
It was during my visit in York in 2011 when I told Paul results that I had found with my friends  Alessandro Toigo and Claudio Carmeli on incompatibility of two observables related to mutually unbiased bases, later reported in \cite{CaHeTo12}.
We then realized that the bound on the incompatibility of such pairs (triangle shaped area (21) in \cite[Prop. 6]{CaHeTo12}) has a natural explanation, valid actually in all operational theories \cite{BuHeScSt13}.
We also had a conjecture that the canonical position and momentum observables are maximally incompatible, but it took some time to prove it \cite{HeScToZi14}.
Remarkably, necessary and sufficient conditions for a pair of maximally incompatible dichotomic measurements to exist in a general probabilistic theory have been found \cite{JePl17}, and those conditions show how the existence links to the geometry of the state space.
  
The earlier convexity argument shows at the same time that the set of all joint observables of a compatible pair of observables is convex. 
It follows that a compatible pair $(\Ao,\Bo)$ has either unique or infinitely many joint observables, the first case urging for further investigation due to its special nature.
One can show that if $\Ao$ (or $\Bo$) is extreme element in the respective set of $m$-outcome ($n$-outcome) observables, then $(\Ao,\Bo)$ has a unique joint observable \cite{HaHePe14}.
Related results have been reported in \cite{GuCu18}.

\section{Witnesses}\label{sec:witnesses}

An \emph{entanglement witness} is a selfadjoint operator $W$ on $\hi^A \otimes \hi^B$ such that $\tr{\omega W} \geq 0$ for all separable states $\omega$ and $\tr{\eta W} < 0$ at least for some entangled state $\eta$.
A physical interpretation is straightforward; one measures an observable that allows to calculate the expectation value $\tr{\omega W}$ (e.g. the spectral measure of $W$) and then directly from that number one concludes that $\omega$ is entangled if $\tr{\omega W}<0$.
A witness measurement cannot verify that $\omega$ is separable as each witness takes positive values also for some entangled states.
The fact that for every entangled state there is a witness that detects it follows from the convexity of the subset of separable states and the standard separation theorems of convex analysis.

The convexity of the subset of all compatible pairs of observables hints that one should be able to detect incompatibility with \emph{incompatibility witnesses}, analogously to entanglement witnesses.
However, as the set of objects is now all pairs of observables with some fixed outcome sets, the physical interpretation is not so straightforward as with bipartite states.
To explain the physically relevant form of incompatibility witnesses, let $\Ao$ and $\Bo$ be two observables with $m$ and $n$ outcomes, respectively.
Fix labeled sets of test states, $\en^{(1)}=\{\varrho^{(1)}_x\}_{x=1}^m$ and $\en^{(2)}=\{\varrho^{(2)}_y\}_{y=1}^n$.
The ability of $\Ao$ to distinguish the states in $\en^{(1)}$ is taken to be the average guessing probability
\begin{equation*}
\pg(\en^{(1)};\Ao) = \tfrac{1}{m} \sum_{x=1}^m \tr{ \varrho^{(1)}_x \Ao(x)}
\end{equation*}
and $\pg(\en^{(2)};\Bo)$ is defined similarly.
The average of these guessing probabilities defines a real-valued function on pairs of observables,
\begin{equation}
\xi(\Ao,\Bo)=\half \pg(\en^{(1)};\Ao) + \half \pg(\en^{(2)};\Bo) \, .
\end{equation} 
The essential property of this function is that 
\begin{equation}
\xi( t (\Ao,\Bo) + (1-t) (\Ao',\Bo')) = t \xi(\Ao,\Bo) + (1-t) \xi(\Ao',\Bo') \, ,
\end{equation}
which is the basic mathematical requirement for a witness.
It can happen that the test states are chosen in a bad way so that $\xi$ does not detect incompatibility, but apart from that unlucky situation there exists a boundary value $c_\xi$ such that $\xi(\Ao,\Bo)\leq c_\xi$ for all compatible pairs $(\Ao,\Bo)$ while $\xi(\Ao',\Bo') > c_\xi$ at least for some incompatible pair $(\Ao',\Bo')$. 
The boundary value has an interpretation as the best guessing probability in the corresponding discrimination task with post-measurement information \cite{CaHeTo18}.
Importantly, if we vary the test states and their prior probabilities, then these type of witnesses are enough to detect the incompatibility of any incompatible pair of observables \cite{CaHeTo19}.
Related aspects of incompatibility detection have been recently developed and discussed in \cite{SkSuCa19,UoKrShYuGu19}.

As an example, let us consider the qubit case and choose $\en^{(1)}=\{ \half (\id \pm \sigma_1)\}$ and $\en^{(2)}=\{ \half (\id \pm \sigma_2)\}$.
We then have
\begin{equation}\label{eq:wit-qubit}
\xi(\Mo^{\alpha,\va},\Mo^{\beta,\vb})= \half ( 1 + \half (a_1 + b_2) )\, .
\end{equation} 
It can be shown that for these collections of states we have 
\begin{equation}\label{eq:wit-bound}
\xi(\Ao,\Bo) \leq \half ( 1+ \tfrac{1}{\sqrt{2}})
\end{equation}
 for any pair of compatible dichotomic qubit observables \cite{CaHeTo18}.
 We thus see that the defined witness detects some incompatible pairs, but a comparison to \eqref{eq:paul} and \eqref{eq:compatible} shows that not all are detected. 
The bound \eqref{eq:wit-bound} corresponds to a scenario where Alice sends a message in a perfectly distinguishable form (i.e. both sets consist of orthogonal pure states), but Bob needs to perform a measurement before Alice announces which was the used encoding (i.e. $\sigma_1$-eigenbases or $\sigma_2$-eigenbases).
If Bob would know the used encoding before his measurement, then he could use an incompatible pair and we see from \eqref{eq:wit-qubit} that he can then even distinguish the states perfectly, as expected. 
The values above the bound, i.e., $a_1 + b_2 > \sqrt{2}$, correspond to choices of measurements that are `too lucky' to be made without knowing the encoding.

\section{Simple condition}\label{sec:ansatz}

Let $\omega$ be a bipartite state.
The partial states of $\omega$ are defined as
$\omega_{[1]}=\ptr{2}{\omega}$ and $\omega_{[2]}=\ptr{1}{\omega}$.
In general, it is not possible to conclude entanglement or separability of $\omega$ from these partial states, but they do however give relevant information that can be used as a quick test of separability.
Firstly, the product state $\omega_{[1]} \otimes \omega_{[2]}$ is different to the original state $\omega$ unless the latter is already a product state.
It is also true that if $\omega_{[1]}$ or $\omega_{[2]}$ is pure, then $\omega$ is separable. 
Further, for a pure state $\omega$ this is a necessary and sufficient condition for separability.
In conclusion, by calculating the partial states one gets an easy sufficient condition for separability, which is also necessary for a special class of bipartite states. 

There exists also a simple condition that implies compatibility and is further necessary for compatibility for a specific type of observables.
Namely, suppose that $\Ao$ and $\Bo$ commute, i.e., $[\Ao(x),\Bo(y)]=0$ for all $x,y$.
Then $\Ao$ and $\Bo$ are compatible as the product of two commuting positive operators is positive and hence $\Go(x,y)=\Ao(x)\Bo(y)$ is a valid joint observable of $\Ao$ and $\Bo$.
Further, if $\Ao$ or $\Bo$ is sharp (i.e., consists of projection operators), then the commutativity is also necessary for $\Ao$ and $\Bo$ being compatible.
In this case the product joint observable is their unique joint observable \cite{HeReSt08}.

There is a also more effective sufficient condition for compatibility than commutativity, but which is still easy to check.
In \cite{BuHe08} I found with Paul that a joint observable for qubit observables $\Mo^{\va}$ and $\Mo^{\vb}$, assuming they are compatible, can be always chosen as
\begin{equation}
\Go(x,y) = \half ( \Mo^{\va}(x)\Mo^{\vb}(y) + \Mo^{\vb}(y)\Mo^{\va}(x) ) \equiv \half \Mo^{\va}(x) \circ \Mo^{\vb}(y)   \, .
\end{equation}
This motivates the following joint observable ansatz \cite{Heinosaari13}.
For any two observables $\Ao$ and $\Bo$, define
\begin{equation}
J_{\Ao,\Bo}(x,y) = \half \Ao(x)\circ \Bo(y) \, .
\end{equation}
The map $J_{\Ao,\Bo}$ satisfies $\sum_y J_{\Ao,\Bo}(x,y) = \Ao(x)$ and $\sum_x J_{\Ao,\Bo}(x,y) = \Bo(y)$, but the operators  $J_{\Ao,\Bo}(x,y)$ are not necessarily positive.
We can take the positivity as a sufficient condition for compatibility:
if $J_{\Ao,\Bo}(x,y) \geq 0$ for all $x,y$, then $\Ao$ and $\Bo$ are compatible.
We call this \emph{Jordan criterion} for compatibility since it uses the Jordan products of operators.  
Clearly, for a commuting pair the Jordan criterion holds, but it also covers many more cases.
Namely, as said before, for a pair of unbiased dichotomic qubit observables the Jordan criterion gives \eqref{eq:paul} and is thus not only sufficient but also necessary.
For biased dichotomic qubit observables the Jordan criterion does not cover all compatible pairs \cite{Heinosaari13}.
However, testing several compatibility criteria with random pairs of observables shows that the  Jordan criterion performs well \cite{HeJiNe19}.

\section{Entanglement/Incompatibility breaking channels}\label{sec:ibc}

Both entanglement and incompatibility are properties that are destroyed under the effect of noise. Paul investigated the influence of noise for joint measurements in several works.
The concept of \emph{unsharp reality} evolved from these investigations and is clearly explained already in \cite{Busch86}.
A comprehensive investigation on the philosophical consequences of the concept is presented in \cite{BuJa10}.

In the extreme cases entanglement and incompatibility can be erased completely. 
Let us first recall that a transformation on a quantum system is describe by a quantum channel, which is a trace preserving completely positive map.
A quantum channel $\Lambda$ is called \emph{entanglement breaking} if $(\Lambda \otimes id)(\omega)$ is separable for all bipartite states $\omega$. 
It is known \cite{HoShRu03} that this is the case if and only if there exists an observable $\Mo$ and states $\varrho_x$ such that
\begin{equation}\label{eq:ebc}
\Lambda(\varrho)=\sum_x \tr{\varrho \Mo(x)} \varrho_x \, .
\end{equation}
Hence, entanglement breaking channels are exactly measure-and-prepare channels.

A channel $\Lambda$ can be equally well consider in the Heisenberg picture, and we denote this map by $\Lambda^*$.
For every observable $\Ao$, the transformed observable $\Lambda^*(\Ao)$ is defined as $\Lambda^*(\Ao)(x) = \Lambda^*(\Ao(x))$.
As in \cite{HeKiReSc15}, we say that a channel $\Lambda$ is \emph{incompatibility breaking} if observables $\Lambda^*(\Ao_1),\ldots,\Lambda^*(\Ao_r)$ are compatible for any choice of observables $\Ao_1,\ldots,\Ao_r$.
A channel $\Lambda$ having the form \eqref{eq:ebc} reads in the Heisenberg picture as
\begin{equation}\label{eq:ebc-h}
\Lambda^*(T)=\sum_x \tr{\varrho_x T} \Mo(x) \, ,
\end{equation}
and we thus see that any transformed observable $\Lambda^*(\Ao)$ is a post-processing of $\Mo$, implying that $\Lambda^*$ is incompatibility breaking.
We conclude that all entanglement breaking channels are incompatibility breaking.
Curiously, there are incompatibility breaking channels that are not entanglement breaking \cite{HeKiReSc15}.
In that sense, incompatibility is more sensitive to noise than entanglement.
It appears to be a difficult problem to characterize all incompatibility breaking channels, although in the Gaussian framework they do have a neat form \cite{HeKiSc15}.

As a side remark, all incompatibility breaking channels are entanglement breaking if we require them to break not only incompatibility of observables but also incompatibility of channels \cite{HeMi17}. An open question is to characterize the subset of devices that determine incompatibility in the sense that their incompatibility is broken only by  measure-and-prepare channels.

\section{More stringent forms of separability/compatibility}\label{sec:classical}

A separable state $\omega$ has, by definition, a convex decomposition of the type \eqref{eq:sep}.
One can make a finer separation of separable states by looking for more specific form of these kind of decompositions.
In particular, if the states $\varrho^A_i$ (or $\varrho^B_i$) can be chosen to be orthogonal pure states, then $\omega$ is said to have \emph{zero discord}.
Separable states with nonzero discord have been shown to have an advantage over zero discord states in some tasks \cite{CaAoBoMoPiWi11}.
We have still a smaller subset of separable states if we require that both collections $\{\varrho^A_i\}$ and $\{\varrho^B_i\}$ can be chosen to be orthogonal pure states; these kind of states are sometimes called \emph{fully classical states}.

By the definition, a compatible pair $(\Ao,\Bo)$ has a joint observable.
Analogously to separable states, we may thus ask if there is a joint observable of some specific type. 
One such special type corresponds to the possibility of measuring $\Ao$ without disturbing $\Bo$, and in this way performing their joint measurment.
The \emph{nondisturbance} means that there is a joint observable $\Go$ of the form
\begin{equation}
\Go(x,y) = \Ii_x(\Bo(y)) \, ,
\end{equation}
where $\Ii$ is an instrument describing some measurement of $\Ao$.
Mathematically, each $\Ii_x$ is a completely positive map such that $\Ii_x(\id)=\Ao(x)$.
The nondisturbance is not a symmetric relation \cite{HeWo10}; it can happen that $\Ao$ can be measured without disturbing $\Bo$ but not vice versa.
For this reason, we have also a smaller class of \emph{mutually nondisturbing pairs}.
For qubit observables both nondisturbance and mutual nondisturbance are equivalent to commutativity \cite{HeWo10}, but in higher dimensions we hence have these tighter forms of compatibility. 
A special form of nondisturbance gives also a physical meaning for the commutativity; as proven by Paul and Javed Singh in \cite{BuSi98}, two observables $\Ao$ and $\Bo$ commute if and only if the L\"uders measurement of $\Ao$ (equivalenty that of $\Bo$) does not disturb $\Bo$.  

The nondisturbance condition gives a physically well motivated criteria to divide compatible pairs into finer classes, but this is not all.
For instance, one can take \emph{broadcastability} still as a more tighter form of compatibility \cite{Heinosaari16}, or one can take a different approach and analyze the minimal number of nonzero elements in all joint observables of a given compatible pair \cite{SkHoSaLi19}.

\section{More than two systems/observables}\label{sec:multi}

Multipartite entanglement refers to entanglement of more than two systems.
The main interests in multipartite entangelement are those features that are not present in bipartite entanglement.
One such phenomenon is the following.
It is possible to have a tripartite entangled state $\omega$ such that all of its three bipartite partial states $\omega_{[12]}=\ptr{3}{\omega}$, $\omega_{[13]}=\ptr{2}{\omega}$ and $\omega_{[23]}=\ptr{1}{\omega}$ are separable. 
In this sense, entanglement is a property of the full state but not seen in any of its parts.
A well-known example of this kind of state is the \emph{GHZ-state} of three qubits, 
\begin{align*}
\omega = \frac{1}{\sqrt{2}} \left( \ket{000} + \ket{111} \right) \, ,
\end{align*}
for which we have
\begin{align*}
\omega_{[12]}=\omega_{[13]}=\omega_{[23]}=  \frac{1}{2} \left( \kb{00}{00} + \kb{11}{11} \right) \, .
\end{align*}

An analogy of the previous phenomena for incompatibility has been discussed e.g. in \cite{LiSpWi11}.
We say that a set $\{ \Ao_1,\ldots,\Ao_p\}$, $p \geq 3$, is a \emph{p-Specker set} if $\Ao_1,\ldots,\Ao_p$ are incompatible but every proper subset is compatible. 
A subset of a compatible set is compatible, hence it is enough to verify that all subsets containing $p-1$ observables are compatible. 

As an example, the triplet of unbiased qubit observables $\Mo^{t\vx}$, $\Mo^{t\vy}$ and $\Mo^{t\vz}$ forms a $3$-Specker set for all values $1/\sqrt{3} < t \leq 1 \sqrt{2}$ \cite{HeReSt08}.
(As an interesting side remark, it has been shown in \cite{QuBuWoCaCa19} that this kind of genuine triplewise incompatibility, i.e., incompatibility that does not reduce to pairwise relations, can be tested in a device-independent way.)
By using generators of a Clifford algebra it was shown that a $p$-Specker set exists for every $p\geq 3$ \cite{KuHeFr14}. 
The minimal Hilbert space dimension in that construction is $2^{p/2}$ for even $p$ and $2^{(p-1)/2}$ for odd $p$.
The Clifford algebra construction, however, does not give the absolute minimal dimension: a $4$-Specker set can be formed by choosing suitable unbiased dichotomic qubit observables \cite{UoLuMoHe16}. 
A recent investigation, reported in \cite{AnKu20}, goes deep into these questions and develops new techniques; it is proven that every $p$-Specker set exists already for qubit measurements.
Specker sets are specific types of \emph{joint measurability structures} and there are still many open questions on this topic; some open questions are listed in \cite{AnKu20}.

\section{Conclusions}\label{sec:conc}

In this paper I have described some analogies between quantum incompatibility and entanglement.
The plurality and attractiveness of different aspects of quantum incompatibility hopefully convinces one to speak about \emph{theory of quantum incompatibility}.
For me, this theory carries the memory of Paul.

\section{Acknowledgement}

I'm grateful to Claudio Carmeli, Ion Nechita, Marco Piani, Alessandro Toigo and Mario Ziman for many valuable discussions and their useful comments on an earlier version of this paper.

This work was performed as part of the Academy of Finland Centre of Excellence program, Project 312058. Financial support from the Academy of Finland, Project 287750, is also acknowledged.

\end{document}